\documentclass[12pt]{article}
\usepackage{amsfonts}

\begin{document}

\author{A. de Souza Dutra$^{a,b}$\thanks{%
E-mail: dutra@feg.unesp.br} and M. Hott$^{b}$\thanks{%
E-mail: hott@feg.unesp.br} \\
$^{a}$Abdus Salam ICTP, Strada Costiera 11, Trieste, I-34100 Italy.\\
$^{b}$UNESP-Campus de Guaratinguet\'{a}-DFQ\thanks{ Permanent
Institution}\\
Departamento de F\'{\i}sica e Qu\'{\i}mica\\
Caixa Postal 205\\
12516-410 Guaratinguet\'{a} SP Brasil}
\title{{\LARGE Dirac equation exact solutions for generalized asymmetrical
Hartmann potentials}}
\maketitle

\begin{abstract}
In this work we solve the Dirac equation by constructing the exact bound
state solutions for a mixing of generalized vector and scalar Hartmann
potentials. This is done provided the vector and scalar potentials hold some
relation. Namely, one must be equals to or minus the other. Finally the case
of some quasi-exactly solvable potentials are briefly commented.

\end{abstract}

\newpage

Very recently in this journal C. Y. Chen \cite{yuan} presented exact
solutions for fermions in the presence of a classical background which is a
mixing of the time-component of a gauge potential and a scalar potential. He
has considered the particular case where both potentials are of the Hartmann
type potential \cite{hartmann}. In fact, he in collaboration with other
authors has dedicated great attention to this mater in the recent years,
which can be verified from a number of interesting works \cite{yuan2}-\cite%
{yuan5}. On the other hand, many years ago, Hautot \cite{hautot} solved the
Schr\"{o}dinger equation for the Coulomb and harmonic oscillator potentials
with some asymmetrical terms like $\frac{f\left( \theta \right) }{r^{2}}$
added. Here we intend to show that one can find some other cases for which
the Dirac equation with classical potentials of vector and scalar natures
with spherical asymmetry can be solved exactly. We begin our study treating
the case of the Hartmann-type and after that we discuss a generalization of
Morse-like potentials \cite{alhaidari}-\cite{castro} with spherical
asymmetry. Finally we comment on the quasi-exactly solvable potentials \cite%
{ces1}-\cite{ces2}.

The generalized Hartmann potential is defined here as
\begin{equation}
V\left( r,\theta \right) =-\,\frac{1}{2}\left( \frac{V_{0}\,\lambda }{r}%
-\hbar ^{2}c^{2}\,\frac{f\left( \theta \right) }{r^{2}}\right) ,  \label{1}
\end{equation}%
\noindent The time-independent Dirac equation for arbitrary scalar and
vector potentials looks like
\begin{equation}
\left[ c\,\vec{\alpha}\cdot \vec{P}\mathbf{+}\mathbb{\beta }\left(
M\,c^{2}+S\left( \vec{r}\right) \right) \right] \psi \left( \vec{r}\right) =%
\left[ E-V\left( \vec{r}\right) \right] \psi \left( \vec{r}\right) ,
\label{2}
\end{equation}

\noindent where it is defined that
\begin{equation}
\vec{P}\equiv -i\,\hbar \,\vec{\nabla},\,\vec{\alpha}\equiv \left(
\begin{array}{ll}
0 & \vec{\sigma} \\
\vec{\sigma} & 0%
\end{array}%
\right) ,\beta \equiv \left(
\begin{array}{ll}
I & 0 \\
0 & -I%
\end{array}%
\right) ,  \label{3}
\end{equation}

\noindent with $\vec{\sigma}$ the vector Pauli spin matrix and $I$ the
identity matrix. Now, using the Pauli-Dirac representation, with
\begin{equation}
\psi \left( \vec{r}\right) =\left(
\begin{array}{l}
\varphi \left( \vec{r}\right) \\
\chi \left( \vec{r}\right)%
\end{array}%
\right) ,  \label{4}
\end{equation}

\noindent we get the following set of coupled equations for the spinor
components
\begin{eqnarray}
c\,\vec{\sigma}\cdot \vec{P}\,\chi \left( \vec{r}\right) &=&\left[ E-V\left(
\vec{r}\right) -M\,c^{2}-S\left( \vec{r}\right) \right] \varphi \left( \vec{r%
}\right) ,  \nonumber \\
&&  \label{5} \\
c\,\vec{\sigma}\cdot \vec{P}\,\varphi \left( \vec{r}\right) &=&\left[
E-V\left( \vec{r}\right) +M\,c^{2}+S\left( \vec{r}\right) \right] \chi
\left( \vec{r}\right) .  \nonumber
\end{eqnarray}

At this point we can treat two non equivalent exact situations. The first
when $S\left( \vec{r}\right) =V\left( \vec{r}\right) $, which was the one
considered in \cite{yuan}, and another where $S\left( \vec{r}\right)
=-\,V\left( \vec{r}\right) $. Once the treatment is quite similar in both
cases, we start by dealing with the first one and then comment about the
second case.

The case with $S\left( \vec{r}\right) =V\left( \vec{r}\right) $, allow us to
decouple the Dirac equation as%
\begin{equation}
\chi \left( \vec{r}\right) =\left[ \frac{c\,\vec{\sigma}\cdot \vec{P}\,}{%
E+M\,c^{2}}\right] \varphi \left( \vec{r}\right) ,  \label{6a}
\end{equation}%
\begin{equation}
\left[ c^{2}\vec{P}\,^{2}+2\,\left( E+M\,c^{2}\right) V\left( \vec{r}\right) %
\right] \varphi \left( \vec{r}\right) =\left[ E^{2}-M^{2}c^{4}\right]
\varphi \left( \vec{r}\right) ,  \label{6}
\end{equation}

\noindent leading us to the following Schr\"{o}dinger like equation
\begin{equation}
\left[ -\hbar ^{2}c^{2}\vec{\nabla}^{2}-\,\left( E+M\,c^{2}\right) \,\left(
\frac{V_{0}\,\lambda }{r}-\frac{f\left( \theta \right) }{r^{2}}\right) %
\right] \varphi \left( \vec{r}\right) =(E^{2}-M^{2}c^{4})\,\varphi \left(
\vec{r}\right) .  \label{7}
\end{equation}

\noindent Performing now the usual separation of variables in spherical
coordinates
\begin{equation}
\varphi \left( \vec{r}\right) =\frac{\,e^{i\,m\,\phi }}{\sqrt{2\,\pi }}\,%
\frac{u\left( r\right) }{r}\,\Theta \left( \theta \right) \,,\,\,\,\,m\in Z,
\label{8}
\end{equation}

\noindent we obtain the equations for $u\left( r\right) $ and $\,\Theta
\left( \theta \right) $, which are respectively given by
\begin{equation}
\frac{1}{\sin \left( \theta \right) }\frac{d}{d\theta }\left( \sin \theta
\,\,\frac{d\,\Theta \left( \theta \right) }{d\theta }\right) -\left[ \frac{%
m^{2}}{\sin ^{2}\theta }+(E+M\,c^{2})\,f\left( \theta \right) -s\right]
\,\Theta \left( \theta \right) =0,  \label{teta}
\end{equation}

\noindent and
\begin{equation}
-\frac{d^{2}u\left( r\right) }{dr^{2}}+\left[ \frac{s}{r^{2}}-\frac{%
\,E+M\,c^{2}}{\hbar ^{2}c^{2}}\,\frac{V_{0}\,\lambda }{r}\right] u\left(
r\right) =\frac{\left( E^{2}-M\,^{2}\,c^{4}\right) }{\hbar ^{2}c^{2}}%
\,u\left( r\right) .  \label{9}
\end{equation}

From the above equation one can see that bound-state solutions are
possible only if $\mid E\mid $ $<Mc^{2}$ and that there is no room
for bound-states for $s<-{1/}{4}$, because this is the critical
value of this parameter, due to the fact that below this value,
the singularity of the potential produces the so called fall to
the center. Finally, regarding the product of parameters
$V_{0}\,\lambda $, there is only one situation where there are no
bound states. That happens if $V_{0}\,\lambda <0$ and $s>0$
simultaneously.

Now, we are able to specify the form of the function $f\left( \theta \right)
$, in order to guarantee that (\ref{teta}) can be exactly solvable.
Following the original work of Hautot \cite{hautot}, we can consider the
cases where
\begin{eqnarray}
f_{1}\left( \theta \right) &=&\frac{\left( \gamma +\beta \,\cos \theta
+\alpha \,\cos ^{2}\theta \right) }{\sin ^{2}\theta };  \label{f1} \\
f_{2}\left( \theta \right) &=&\frac{\left( \gamma +\beta \,\cos ^{2}\theta
+\alpha \,\cos ^{4}\theta \right) }{\sin ^{2}\theta \cos ^{2}\theta };
\label{f2} \\
f_{3}\left( \theta \right) &=&\gamma +\beta \,\cot \theta +\alpha \,\cot
^{2}\theta .  \label{f3}
\end{eqnarray}

For each one of the above configurations for $f(\theta )$, Hautot was able
to map equation (\ref{teta}) into a hypergeometric differential equation,
which has finite solutions for $\Theta \left( \theta \right) $ in the range $%
0\leq \theta \leq \pi $ provided that $s$ satisfies some constraint as a
function of the parameters $\alpha $, $\beta $, $\gamma $ and a new quantum
number $k=0,1,2,...$, that was introduced in order to render $\Theta \left(
\theta \right) $ a finite polynomial (Jacobi polynomial). Here we present a
detailed analysis of the first case and comment about the other cases.

For $f_{1}(\theta )$ and $f_{2}(\theta )$ the regular solutions found for $%
\Theta (\theta )$ are generally given by

\begin{equation}
\Theta _{k}(\theta )=z^{\rho }(1-z)^{\nu }\,_{1}F_{2}(-k,b,d;z),  \label{10}
\end{equation}

\noindent where $_{1}F_{2}(-k,b,d;z)$ is the hypergeometric function and $%
z=\cos ^{2}\theta $

For $f_{1}(\theta )$ one has%
\begin{eqnarray}
\rho &=&\frac{1}{2}\left[ m^{2}+(E+Mc^{2})(\alpha -\beta +\gamma )\right]
^{1/2},  \nonumber \\
\nu &=&\frac{1}{2}\left[ m^{2}+(E+Mc^{2})(\alpha +\beta +\gamma )\right]
^{1/2},  \label{11a}
\end{eqnarray}

\begin{eqnarray}
b &=&k+2(\rho +\nu )+1,  \nonumber \\
d &=&1+2\rho ,  \label{11}
\end{eqnarray}

and

\begin{equation}
s+\frac{1}{4}=\frac{1}{4}(b+k)^{2}-(E+Mc^{2})\alpha .  \label{12}
\end{equation}

At this point some discussion on the range of validity of the
potential parameters should be done. Note that in order to keep
$b$ and $d$ real, it is necessary to impose that $\alpha +\gamma
-sign\left( \beta \right), \,\beta \geq 0$. Furthermore, the
imposition that the system should avoids the fall to the center,
implies into the following additional restriction,

\begin{equation}
s_{min}+\frac{1}{4}=\frac{1}{4}(b_{min})^{2}-(E+Mc^{2})\alpha \geq \frac{1}{%
16}.  \label{min}
\end{equation}

Defining now the variables $y \equiv \sqrt{E+Mc^2}$, $\delta_1 \equiv \alpha
-\beta +\gamma$ and $\delta_2 \equiv \alpha +\beta +\gamma$, we obtain the
following equation

\begin{equation}
\left[ \left( \sqrt{\delta_1}+\sqrt{\delta_2} \right)^2-4\, \alpha \right]\,
y^2+2\,\left( \sqrt{\delta_1}+\sqrt{\delta_2} \right) \, y + \frac{3}{4}
\geq 0 .
\end{equation}

On the other hand, we know that $y$ is a positive definite variable. As a
consequence, it is not hard to conclude that the only way to avoid any
further restrictions over the energy of the system, is to require that the
coefficient of $y^{2}$ be positive definite also. This requirement implies
into the following equation for the parameters

\begin{equation}
\alpha +\gamma +\sqrt{\left( \alpha +\gamma \right) ^{2}+\beta ^{2}}%
-4\,\alpha \geq 0.
\end{equation}%
\smallskip

\smallskip

A very similar analysis could be done for $f_{2}(\theta )$, because in this
case we have%
\begin{eqnarray}
\rho &=&\frac{1}{4}+\frac{1}{4}\left[ 1+4(E+Mc^{2})\gamma \right] ^{1/2},
\nonumber \\
\nu &=&\frac{1}{2}\left[ m^{2}+(E+Mc^{2})(\alpha +\beta +\gamma )\right]
^{1/2},  \label{13a}
\end{eqnarray}

\begin{eqnarray}
b &=&k+2(\rho +\nu )+\frac{1}{2},  \nonumber \\
d &=&2\rho +\frac{1}{2},  \label{13}
\end{eqnarray}

and

\begin{equation}
s+\frac{1}{4}=(b+k)^{2}-(E+Mc^{2})\alpha .  \label{14}
\end{equation}

Its analysis is straightforward and we let it for the interested reader.
Althought the case with $f_{3}(\theta )$ needs a different change of
variables in order to cast the equation (\ref{teta}) into a familiar
hypergeometric one, it presents the same general behavior in terms of
restriction over the potential parameters, and we do not present it here
also.

It is worth to mention that our analysis could be extended by
including some other cases for $f(\theta )$, as was done in
reference \cite{hautot}, had we taken into account a
two-dimensional scenario.\medskip

The solution for the radial equation (\ref{9}), by its turn, can
be obtained exactly following the reference \cite{merz}. It
corresponds to an effective
radial equation when a non-relativistic particle of effective mass $%
M_{eff}=1/2$ is under the action of an effective potential

\begin{equation}
V_{eff}=\frac{\hbar ^{2}s}{r^{2}}-\frac{\,E+M\,c^{2}}{c^{2}}\,\frac{%
V_{0}\,\lambda }{r}  \label{15}
\end{equation}

\noindent and effective energy given by

\begin{equation}
E_{eff}=\frac{E^{2}-M\,^{2}\,c^{4}}{c^{2}}.  \label{16}
\end{equation}

The normalized radial eigenfunctions $u(r)$ can also be read off directly
from reference \cite{merz} and is given by

\begin{equation}
u_{n,l}(r)=\left\{ (2\kappa )^{3}\frac{\Gamma (n+1)}{2n[\Gamma (n+2l+2)]^{3}}%
\right\} ^{1/2}\,\exp (-\kappa r)\,(2\kappa
r)^{l+1}\,\,L_{n-l-1}^{2l+1}(2\kappa r),  \label{17}
\end{equation}

\noindent where $L_{n-l-1}^{2l+1}(2\kappa r)$ are the Laguerre polynomials, $%
n=0,1,2,...$ denotes the number of nodes of the radial function. We have
defined

\begin{equation}
\kappa =\sqrt{\frac{(M^{2}c^{4}-E^{2})}{\hbar ^{2}c^{2}}\,\,}%
\,\,\,\,\,\,\,\,\,\,\mathrm{and\,\,\,\,\,\,\,\,\,\,}l=2\sqrt{s+1/4}-1/2>0.
\label{18}
\end{equation}

The effective energy satisfies the following relation
\begin{equation}
\frac{\left( E^{2}-M\,^{2}\,c^{4}\right) }{c^{2}}=-\left[ \frac{%
(E+Mc^{2})V_{0}\lambda }{2\hbar c(n+l+1)}\right] ^{2},  \label{19}
\end{equation}

\noindent from which we can take the energy eigenvalues. At this point we
could be tempted to write a solution

\begin{equation}
E=Mc^{2}\frac{1-\tau ^{2}}{1+\tau ^{2}},  \label{20}
\end{equation}

\noindent where $\tau =\frac{V_{0}\lambda }{2\hbar c(n+l+1)}$, as was done
in \cite{yuan}. However, we should remind that in this case $l$ is a
nontrivial function of the energy leading to a somewhat intricate equation
for the energy.

\bigskip

Finally one can note that the equations for the bispinors $\varphi (\vec{r})$
and $\chi (\vec{r})$ in the situation where $S\left( \vec{r}\right)
=-\,V\left( \vec{r}\right) $ are given respectively by
\begin{equation}
\varphi \left( \vec{r}\right) =\left[ \frac{c\,\vec{\sigma}\cdot \vec{P}\,}{%
E-M\,c^{2}}\right] \chi \left( \vec{r}\right)   \label{21}
\end{equation}

\noindent and

\begin{equation}
\left[ c^{2}\vec{P}\,^{2}+2\,\left( E-M\,c^{2}\right) V\left( \vec{r}\right) %
\right] \chi \left( \vec{r}\right) =\left[ E^{2}-M^{2}c^{4}\right] \chi
\left( \vec{r}\right) .  \label{22}
\end{equation}

These can also be obtained by performing the transformations

\begin{equation}
\varphi \rightarrow \chi ,~~~~\chi \rightarrow \varphi ,~~~~~V\rightarrow
V~~~~\mathrm{and~~~~}E\rightarrow -E  \label{23}
\end{equation}%
in the equations (\ref{6a}) and (\ref{6}). Then the solutions for the
situation where $S\left( \vec{r}\right) =-\,V\left( \vec{r}\right) $ can be
obtained from those in which $S\left( \vec{r}\right) =\,V\left( \vec{r}%
\right) $ by means of the set of transformations (\ref{23}).

\bigskip

\textbf{Acknowledgments:} The authors are grateful to CNPq for
partial financial support. This work has been finished during a
visit (ASD)\ within the Associate Scheme of the Abdus Salam ICTP.

\end{document}